\documentclass[11pt]{article}
\usepackage{amsmath,times}
\usepackage[english]{babel}
\usepackage[pdftex]{graphicx}
\usepackage{float}
\usepackage{amsfonts}
\usepackage[utf8]{inputenc}
\usepackage[left=3cm,right=2cm,top=2cm,bottom=2cm]{geometry}
\usepackage{subfig}
\usepackage{setspace}
\spacing{1.5}
\title{\bf Dynamics of 1-D Electron Motion under a Periodic Electric Field with Radiation Reaction  Effect }
\author{Gustavo V. L\'opez\footnote{gulopez@cencar.udg.mx}~ and Jorge Lizarraga\footnote{thejorge@live.com.mx}~\\ \\
 Departamento de F\'{i}sica, Universidad de Guadalajara,\\
 Blvd. Marcelino Garc\'{i}a Barragan y Calzada Ol\'{i}mpica, \\44200 Guadalajara, Jalisco, Mexico}
\begin{document}
\maketitle
\renewcommand{\abstractname}{\Large Abstract}
\begin{abstract}
\vskip1pc\noindent
We consider the 1-D motion of an electron under a periodic force and taking into account the effect of radiation reaction dissipation force on its motion, using the  formulation of the radiation reaction force as a function of the external force. Two cases are considered: a simple sinusoidal time depending force, and sinusoidal electromagnetic force with position and time dependence. We found that the difference of the normalized  (with respect the speed of light) velocities, with and without radiation reaction,  are quite small between $10^{-31}$ to $10^{-14}$ for intensities on the electric field of $10^{-8}$ to 1 $Dynes/ues$, which may represent some concern to measure experimentally.    
\end{abstract}
\centerline{\large {\bf PACS:} 41.20.-q, 41.60.-m, 41.60.Ap}
\centerline{\large {\bf Key Words:} Radiation Reaction Force, 1-D electron motion, damping motion}
\newpage
\section{Introduction}
Emission of electromagnetic radiation  is a natural classical phenomenon appearing because of the acceleration of any charged particle~\cite{J1962,landau1971}. This radiation implies an emission of energy of the charge and, consequently, a damping effect on its motion. which is modeled through a force called "radiation reaction force." An alternative formulation to the usual Abraham-Lorentz-Dirac radiation reaction force~\cite{Mabrahanm,Halorentz,dirac1938classical} has appear recently~\cite{lopez2016force1,lopez2016generalization} , where the radiation reaction force is given in terms of the same external force which brings about the acceleration of the charge. This approach intents  to solve some difficulties of the previous formulation~\cite{comay1993remarks,valentini1988resolution,spohn2004dynamics,heras2006preacceleration,
jackson2007comment,hnizdo2007comment,heras2007can,griffiths2010abraham}, but it 
has not been experimentally tested yet, although some theoretical studies has been done already~\cite{lopez2016force1,lopez2016force2}. 
Because of this experimental requirement, we make in this paper a study of the 1-D motion of an electron under a periodic external force. The modified relativistic equation of 1-D motion of a charged particle is given by~\cite{lopez2016force1}  
\begin{equation}\label{pex}
\frac{d(\gamma mv)}{dt}=F-\frac{\lambda_0F^2}{v},\quad\quad \lambda_0=\frac{2q^2}{3m^2c^3}.
\end{equation}   
where $\gamma=(1-\beta^2)^{-1/2}$ with  $\beta=v/c$ being the normalized velocity of the charge, and $F$ is the external force. This is the equation of motion that we will use in our study. We will study first the dynamics under the simple sinusoidal force
\begin{equation}\label{fo1}
F=qE_{0}\cos(\omega t + \phi),
\end{equation}
where $E_0$, $\omega$ and $\phi$ are the amplitude, the angular frequency and the phases of the electric field, and then we will extend our study to the position and time depending force
\begin{equation}\label{fo2}
F=qE_0\cos(kx-\omega t + \phi),\quad\quad k=\omega/c.
\end{equation}
\section{Dynamical Equations}
Using the external force (\ref{fo1}) and (\ref{fo2}) in (\ref{pex}), and after some rearrangements, we get the following dynamical systems
\begin{subequations} 
\begin{eqnarray}
&\dot{\beta}&=\left(q E_{0}\cos(\omega t + \phi)-\frac{q^{2}\lambda_{0}E_{0}^{2}}{\beta c}\cos^{2}(\omega t + \phi)\right)\frac{(1-\beta^{2})^{3/2}}{mc}\label{beta_punto_1}\label{eqa1}\\
&\dot{x}&=\beta c \label{equis_punto_1}\label{eqa2}
\end{eqnarray}
\end{subequations}
and
\begin{subequations} 
\begin{eqnarray}
&\dot{\beta}&=\left(q E_{0}\cos(kx-\omega t + \phi)-\frac{q^{2}\lambda_{0}E_{0}^{2}}{\beta c}\cos^{2}(kx-\omega t + \phi)\right)\frac{(1-\beta^{2})^{3/2}}{mc}\label{beta_punto_2}\label{eqb1}\\
&\dot{x}&=\beta c \label{equis_punto_2}\label{eqb2}
\end{eqnarray}
\end{subequations}
In the first dynamical system the motion in the coordinates "$x$" and "$\beta$" are disconnected, but on the second dynamical system these coordinates are coupled. These dynamical systems are solved using Runge-Kutta method at fourth order and are solved for the cases without radiation reaction force ($\lambda_0=0$, the solution is denoted as $\beta_0(t)$) and with radiation reaction force 
($\lambda_0\not=0$, the solution is denoted as $\beta(t)$). Then, we calculate the difference
\begin{equation}\label{diff}
\Delta\beta=\beta(t)-\beta_0(t).
\end{equation}
\section{Numerical Analysis}
\begin{figure}[H]
{\centering
 \includegraphics[width=0.85\textwidth]{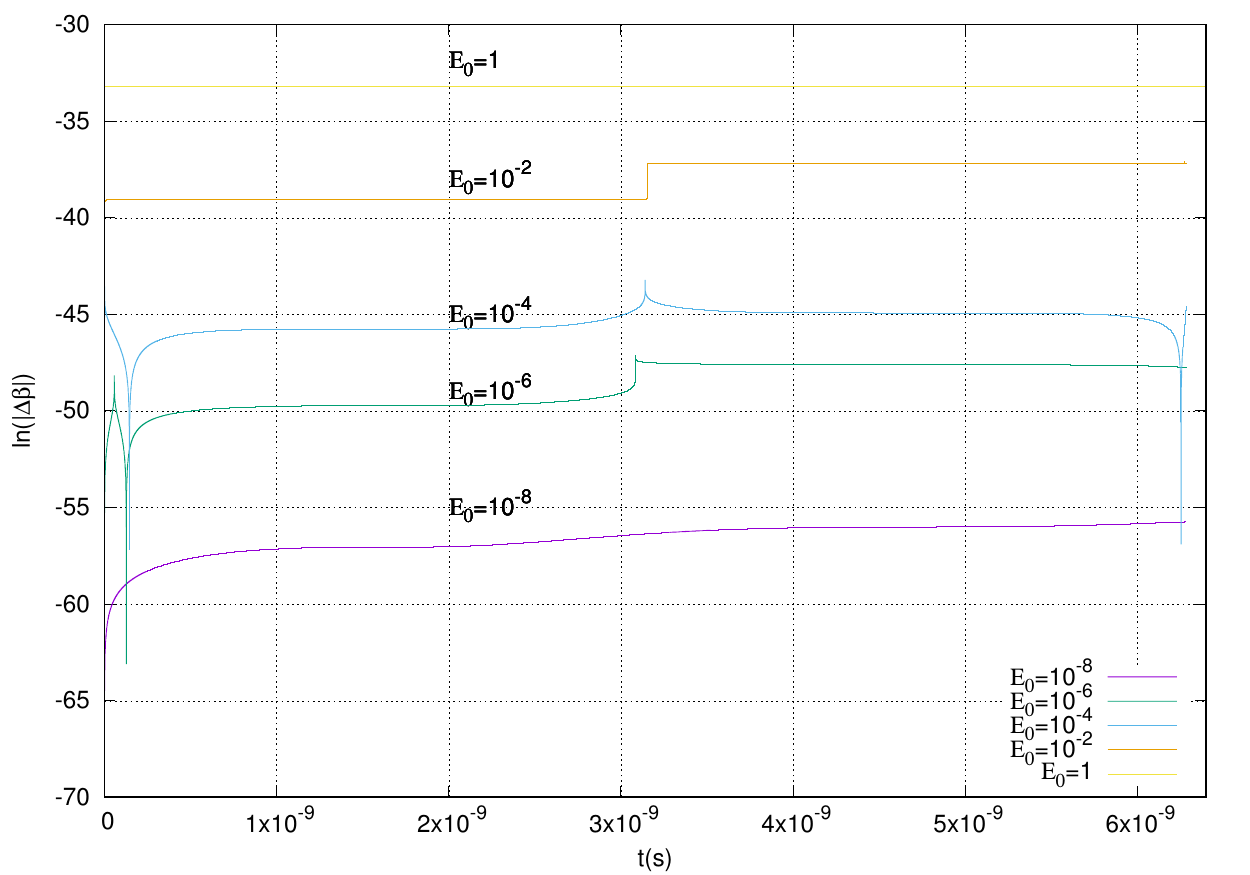}
 \caption{Dynamical system (4): Logarithm of $|\Delta\beta|$ as a function of time for different  electric field intensities in units of Dynes/ues with a fixed frecuency of $\omega=1\times 10^{9}$.}
 \label{Variacion_Campo_Electrico}}
 \end{figure}

\begin{figure}[H]
{\centering
 \includegraphics[width=0.85\textwidth]{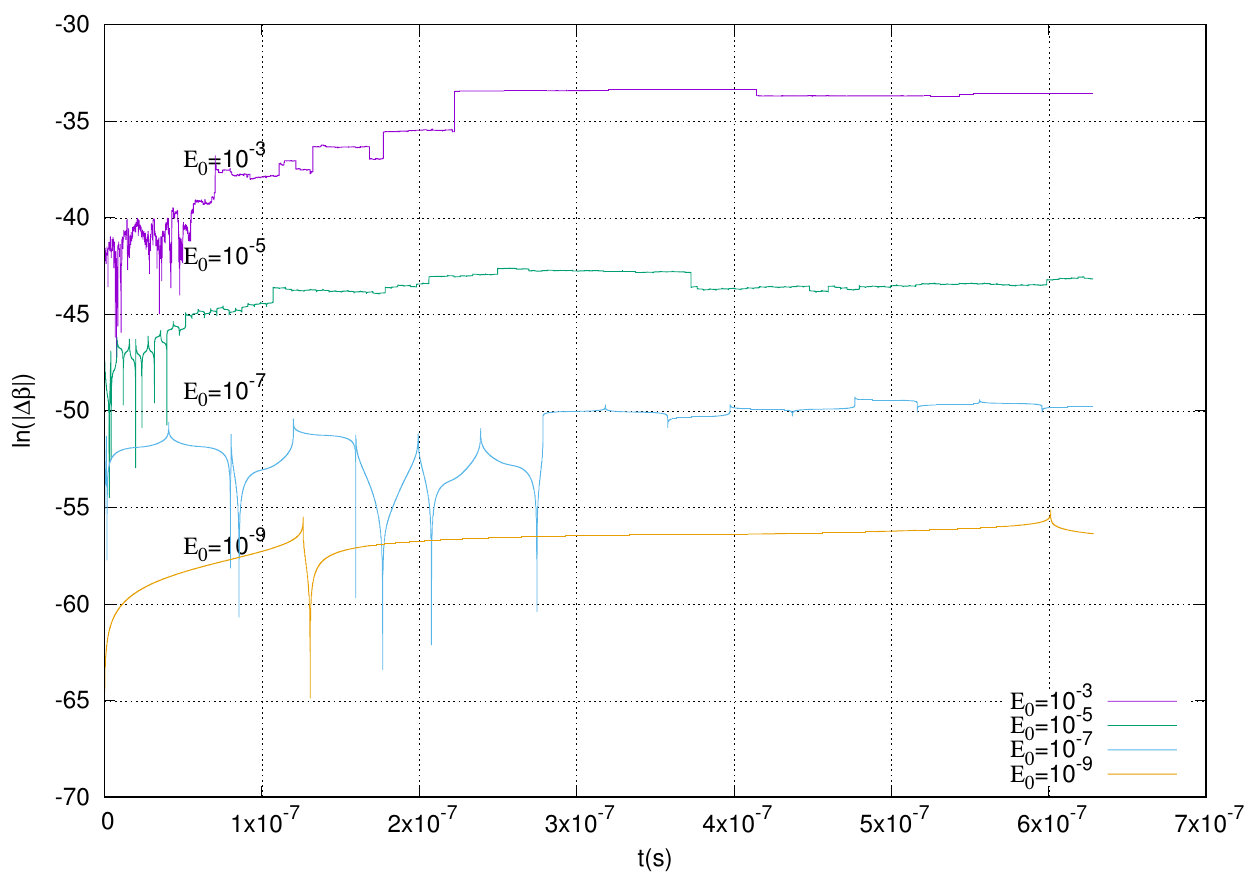}
 \caption{Dynamical system (5):  Logarithm of $|\Delta\beta|$ as a function of time  for fixed frequency $\omega=1\times 10~GHz$ and several $E_0$ values. }
 \label{Variacion_Campo_Electrico_dependiete_posicionYtiempo}}
 \end{figure}

\begin{figure}[H]
{ \centering
    \includegraphics[width=0.85\textwidth]{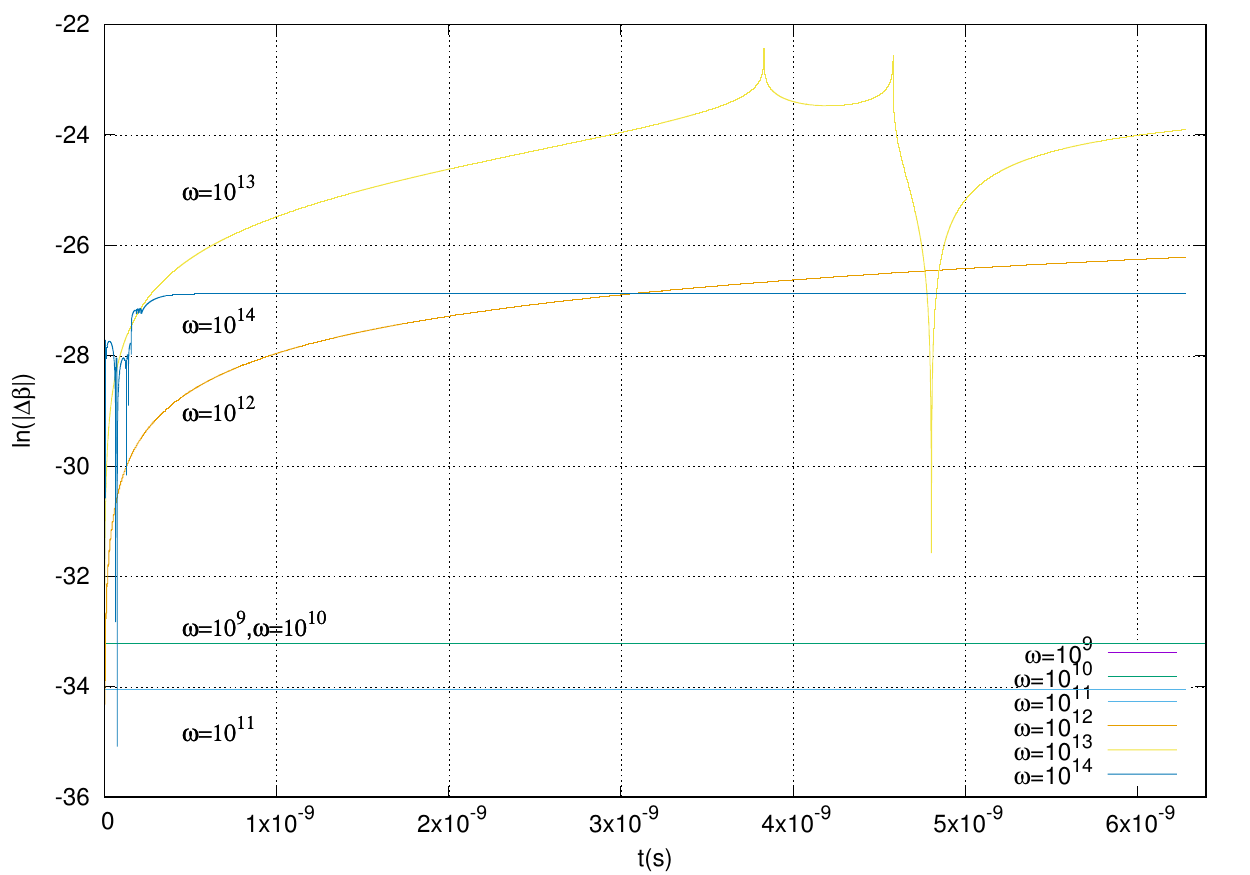}
   \caption{Dynamical system (4):  Logarithm of $|\Delta\beta|$ as a function of time for different frequencies (in $Hz$) and fixed $E_0=1~Dynes/ues$. }
 \label{frecuencias1}}
\end{figure}

We solved the first dynamical system, equations~\ref{beta_punto_1} and~\ref{equis_punto_1}, for different but fixed $E_{0}$, electric field intensities, to see the change in the electron's behavior. The Figure~\ref{Variacion_Campo_Electrico} shows the expression $\ln(|\Delta\beta)$,  solution of equations (\ref{eqa1}) and (\ref{eqa2}), as a function of time for several $E_0(Dynes/ues)$ values. 
As one can see, this figure has some peaks and discontinuity, the one that is in the middle can be associated with a change of force's direction, this will happen when $\omega t=\pi$ (for this case $\omega=1\times 10^{9}$ and $\phi=0$). The peaks at the very beginning are due to the fact that  $\beta$ and $\beta_{0}$ are very close together in magnitude, that is $\Delta \beta\sim 0$ and as is know $ln(0)= -\infty$. 
in addition, we can see that the difference $\Delta\beta$ increases  with the intensity of electric field. From equation~\ref{beta_punto_1} we notice that when the electron's normalized velocity is closed to one ( $\beta\sim 1$),  the external force approaches to zero leading to a $\Delta\beta$ almost constant, and this is observed  for $E_{0}=1$.  
Figure~\ref{Variacion_Campo_Electrico_dependiete_posicionYtiempo} shows the same expression as before but with the solutions for the second dynamical system~(\ref{eqb1}) and~(\ref{eqb2}). For this case, the electron's behavior is a little more complicated, as we can see there are more peaks and discontinuities. As $E_{0}$ increases the difference $\Delta\beta$ increases for  times above $3\times 10^{-7}sec$ where almost a constant value is reached. This is happening since the electron's speed is near the speed the force is close to zero. The peaks appears because the difference between normalized velocities starts to decrease at those points, and this occurs when the $\beta_{0}$ is very close to the speed of light. Thus  from~(\ref{beta_punto_2}), the external force starts to decrease leading to $\Delta\beta\sim 0$.  
Finally we studied the case for different frequencies leaving the electric field intensity fixed at $E_{0}=1 Dynes/ues$. The results for the dynamical system (\ref{eqa1}) and (\ref{eqa2}) 
are shown in Figure~\ref{frecuencias1}.  We can see that for higher frequencies it appears more peaks and discontinuities at the beginning of the motion of the electron, and the reason is due to the same as the previous case,  electron  speed leads to a difference $\Delta\beta\sim 0$. Also we have three cases of interest here, first one for frequencies of order $1GHz$ to $10~GHz$, the second one for $100~GHz$ to $1000~GHz$ and the last one when $\omega\sim 10000~GHz$. For the first case, we observe that $\Delta\beta$ is almost constant, this is because electron's speed reach out the relativistic case very soon and the external force disappears. For the second one, force's frequency leads to a change of force's direction in a way that the electron doesn't reach the relativistic case for the simulation time considered. And for the last case, the frequency makes the force changes direction so rapidly that we start to have a lot of discontinuities until electron speed is stable.\\
\noindent For the dynamical system (5), we have fixed the electric field intensity at $E_{0}=1\times 10^{-7} Dynes/ues$ and we change the frequencies from $\omega=(1,1000)GHz$. One can see that electron's behavior is very complicated, we have more discontinuities while the frecuency is higher until the electron reaches a relativistic speed at frecuency $\omega=1\times 1000~GHz$ and the external force disappears. Those discontinuities can be explained similarly as we have done with all the previous cases. 
 \begin{figure}[H]
{\centering
 \includegraphics[width=0.85\textwidth]{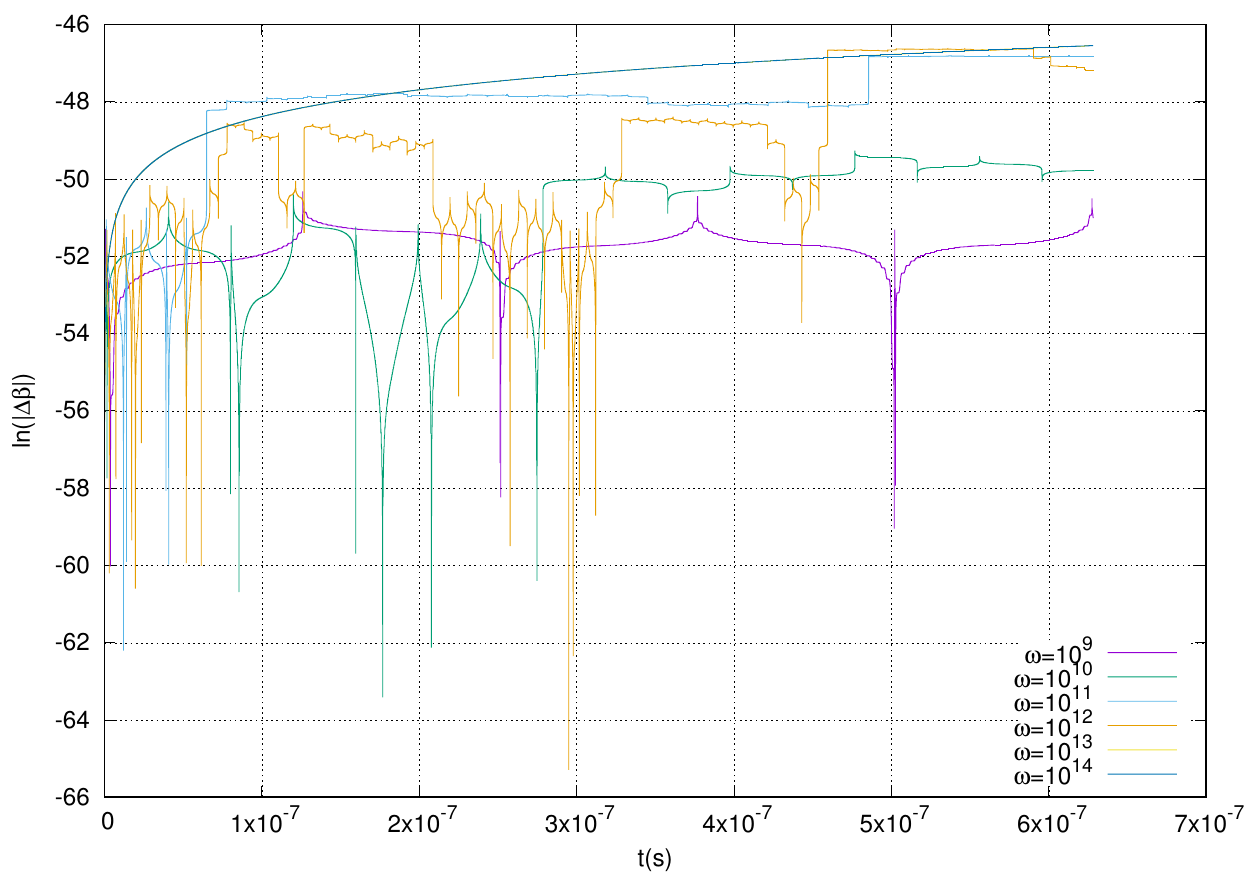}
 \caption{Dynamical system (5):  Logarithm of $|\Delta\beta|$ as a function of time  for fixed $E_0=10^{-7}Dynes/ues$ and several frequencies (in $Hz$). }
 \label{Variacion_Frecuencia_dependiete_posicionYtiempo}}
 \end{figure}
\section{Conclusions}
We studied 1-D electron motion under two periodic forces and including the radiation reaction force. 
When the periodic external force has a simple time dependence (dynamical system (4)), we found that the difference $\Delta\beta$ increases  with $E_{0}$. For the the periodic force depending on position and time (dynamical system (5)), the relation between $\Delta\beta$ and $E_0$  is more complicated but the same dependence is observed after some time
( $3\times 10^{-7} sec$) when the difference becomes more stable. On the other hand and for both dynamical systems, the relation between the frequency of the electric field and $\Delta\beta$ is different since several peaks and discontinuities appears. The observed value for $\Delta\beta$ for both cases is of the order of $ 10^{-31}$ to $10^{-14}$ which could be a concern to detect experimentally, but we think it worths to try to make an experiment to see whether or not this approach for radiation reaction force points to the right experimental direction.
\newpage

\end{document}